\documentclass[a4paper]{article}

\newcommand{\be}{\begin{equation}}
\newcommand{\ee}{\end{equation}}
\newcommand{\g}{{\bf g}}

\newcommand{\A}{{\bf a}}
\newcommand{\C}{{s}}

\newcommand{\T}{\mbox{\bf T}}

\begin{document}
\begin{flushright}
\hfill CAMS/98-02\\
\hfill hep-th/9806181\\
\end{flushright}
\vspace{1cm}
\begin{center}
\baselineskip=16pt
{\Large\bf Gauged N=4 Supergravity through Compactification and BPS
Monopoles}
\vskip 2cm
{\bf Ali H. Chamseddine}\footnote{Plenary talk given at the
PASCOS-98 meeting at Northeastern University, Boston, March 21-29,
1998}\\
\vskip 0.5cm
{\em Center for Advanced Mathematical Sciences\\
and\\
Physics Department\\
American University of Beirut \\
Beirut, Lebanon}\\
E-mail chams@aub.edu.lb
\vskip 0.2cm
\end{center}
\vskip 1cm
\begin{abstract}
It is shown that N=4 gauged supergravity in four dimensions is
obtained by compactifying N=1 supergravity in ten dimensions on the
group manifold $S^3\times S^3$. This could be further related to
supergravity in eleven dimensions.  Analysis of  supersymmetry
conditions of N=4 gauged supergravity in four dimensions reveals
solutions which preserve 1/4 of the supersymmetries and are
characterized by a BPS-monopole-type gauge field.  These solutions
are lifted to solutions of the ten and eleven dimensional theories.
\end{abstract}
\bigskip
\bigskip
\newpage
\section{Introduction}
Gauged supergravities in four dimensions have been known to be
related to compactifications of supergravities in higher
dimensions.  Some of these results have been shown explicitly, but
for others such as the conjectured relation between gauged N=8
supergravity in four dimensions and compactified eleven dimensional
supergravity on $S^7$  a full proof is still missing \cite{DNP}.
For the simpler gauged N=4 supergravity, the no-go theorem of
\cite{FGW} stood as an obstacle to such a realisation. It turns out
that the correct compactification of the ten dimensional theory is
to take the internal six dimensionnal manifold to be the group
manifold $S^3\times S^3 $. This is to be suplemented with a very
specific relation between the components of the metric and the
antisymmetric tensor along the internal directions. I will show how
this is done explicitly, and further relate this to a
compactification of eleven dimensional supergravity. I will also
describe non-abelian monopole solutions preserving $\frac{1}{4}$ of
supersymmetries. These are the only known non-abelian solutions
which are not conformally flat or obtained by embedding the
gravitational spin-connection in the gauge group \cite{DKL}. Using
our explicit relation between ten-dimensional supergravity and the
Friedman-Schwarz model, the four dimensional solution is lifted to
a ten-dimensional one. This could be further lifted to a solution
of eleven-dimensional supergravity. The results presented here  is
based on work done in collaboration with M. Volkov \cite{CV}.
\section{Compactification of ten-dimensional supergravity on $S^3\times S^3$}
It is well known that torus compactification of ten-dimensional
supergravity to four dimensions gives N=4 supergravity coupled to
six vector multiplets \cite{chams80}. The four dimensional vectors
of supergravity and matter are linear combinations of the
components of the metric and antisymmetric tensor along the
internal directions. Truncating the matter multiplets is equivalent
to identifying  some components of the ten-dimensional metric with
those of the antisymmetric tensor. Another known compactification
of ten-dimensional supergravity is due to Scherk and Schwarz
\cite{SS} where the internal space is taken as a group manifold.
The maximal allowed group is $S^3\times S^3$. Such a study was done
in \cite{chams81} for the dual formulation of supergravity. The
resulting N=4 four dimensional supergravity theory has a
non-compact gauge group containing $SU(2)\times SU(2)$. It is not
possible to truncate the resulting theory to give the
Friedman-Schwarz model. A careful analysis of this model suggests
that the antisymmetric tensor must play a  non-trivial role. We
shall now show explicitly the correct compactification.

The bosonic part of  N=1 supergravity action in ten dimensions is:
\begin{equation}
S_{10}=\int \left( -\frac{\hat{e}}{4}\,\hat{R}+\frac{\hat{e}}{2}\,\partial
_{M}\hat{\phi}\,\partial ^{M}\hat{\phi}+\frac{\hat{e}}{12}\,e^{-2\hat{\phi}}%
\hat{H}_{MNP}\,\hat{H}^{MNP}\right) \,d^{4}x\,d^{6}z\equiv S_{\hat{G}}+S_{%
\hat{\phi}}+S_{\hat{H}}.\label{tenaction}
\end{equation}
The notation is as follows:  hatted symbols are used for
10-dimensional quantities. Late capital Latin letters stand for the
base space indices, and  early letters refer to  tangent space
indices. For four-dimensional space-time indices, late and early
Greek letters denote base space and tangent space indices,
respectively. Similarly, the internal base space and tangent space
indices are denoted by late and early Latin letters, respectively:
\begin{equation}
\{M\}=\{\mu =0,\ldots ,3;\,m=1,\ldots ,6\},\ \ \ \{A\}=\{\alpha =0,\ldots
,3;\,{\rm a}=1,\ldots ,6\}.\ \
\end{equation}
The general coordinates $\hat{x}^{M}$ consist of spacetime coordinates $x^{{%
\mu }}$ and internal coordinates $z^{m}$. The flat Lorentz metric of the
tangent space is chosen to be $(+,-,\ldots ,-)$ with the internal dimensions
all spacelike.
One has $\hat{e}=\left| \hat{e}_{\ M}^{A}\right| $, the metric
is related to the vielbein by $\hat{{\bf g}}_{MN}=\hat{\eta}_{AB}\hat{e}%
_{\ M}^{A}\hat{e}_{\ N}^{B}=
\eta _{\alpha \beta }\hat{e}_{\ M}^{\alpha }\hat{e}%
_{\ N}^{\beta }-\delta _{{\rm ab}}
\hat{e}_{\ M}^{{\rm a}}\hat{e}_{\ N}^{{\rm b}}\,,
$ and the antisymmetric tensor field strength is
\begin{equation}
\hat{H}_{MNP}=\partial _{M}\hat{B}_{NP}+\partial _{N}\hat{B}_{PM}+\partial
_{P}\hat{B}_{MN}\,.
\end{equation}
The internal space spanned by $z^{m}$ is assumed to form a compact group
space. This means that there are functions
$U_{\ m}^{{\rm a}}\,(z)$ subject to
the condition
\begin{equation}
\left.\left.
\left( U^{-1}\right) _{{\rm b}}^{\ m}
\left( U^{-1}\right) _{{\rm c}}^{\ n}\right(
\partial _{m}U_{\ n}^{{\rm a}}-\partial _{n}U_{\ m}^{{\rm a}}\right)
=\frac{f_{%
{\rm abc}}}{\sqrt{2}}\, ,
\end{equation}
where $f_{{\rm abc}}$ are the group structure constants. The volume
of the space is $\Omega =\int \left| U_{\ m}^{{\rm a}}\right|
d^{6}z\, $. In the particular case when the internal space is the
product manifold SU(2)$\times $SU(2) it is convenient to
parametrize  the 6 internal coordinates by a pair of indices:
$\{m\}=\{(\C ),i\}$, where ${s}=1,2$ and $i=1,2,3$ and  similarly
for the tangent space coordinates:$%
\{{\rm a}\}=\{(\C),a\}$, $a=1,2,3$. Each of the two $S^{3}$'s admits
invariant 1-forms $\theta ^{(\C)\,a}=\theta _{\ \ \
i}^{(\C)\,a}dz^{(\C)\,i}$ satisfying $ d\theta ^{(\C)\,a}+
\frac{1}{2}\, \epsilon _{abc}\,\theta ^{(\C)\,b}\wedge \theta ^{(\C)%
\,c}=0\,$. If we choose $U_{\ m}^{{\rm a}}\equiv U_{\ \ \ i}^{(\C)\,a}=
-\frac{\sqrt{2}}{g_{{s}}}\,\theta
_{\ \ \ i}^{(\C)\,a}\, $,
where $g_{{s}}$ are the two gauge coupling constants, then the
structure constants  will be $f_{{\rm abc}}\equiv
f_{abc}^{(\C)}=g_{{s}}\,\epsilon _{abc}\, $. Similarly, if one of
the gauge coupling constants vanishes, say $g_{2}=0$, the internal
space is SU(2)$\times \left[ {\rm U}(1)\right] ^{3}$.

{\bf 1. The metric and the dilaton.--} Let us now return to the
general parametrization of the internal space. Compactification of
the action (\ref{tenaction}) starts by choosing the vielbein and
the dilaton in the following form:
\[
\hat{e}_{\ \mu }^{\alpha }\,=e^{-\frac{3}{4}\phi }\,
e_{\ \mu }^{\alpha }\,,\ \ \ \
\ \
\hat{e}_{\ \mu }^{{\rm a}}=\sqrt{2\,}e^{\frac{1}{4}\phi }\,
A_{\mu }^{{\rm a}%
}\,,\ \ \ \ \
\]
\be
\hat{e}_{\ m}^{\alpha }\,=0\,,\ \ \ \ \
\hat{e}_{\ m}^{{\rm a}}=e^{\frac{1}{4}%
\phi }\,U_{\ m}^{{\rm a}}\,,\ \ \ \ \ \hat{\phi}=-\frac{\phi }{2}\,, \label{vier}
\ee
where all quantities on the right, apart from $U_{\ m}^{{\rm a}}$,
depend only
on $x^{\mu }$. One has $\hat{e}=e^{-3\phi /2}\left| U_{\ m}^{{\rm a}}\right|
\,e$. The dual basis is given by
\[
\hat{e}_{\alpha }^{\ \mu }\,=e^{\frac{3}{4}\phi }\,e_{\alpha }^{\ \mu }\,,
\ \ \
\ \ \ \hat{e}_{{\rm a}}^{\ \mu }=0\,,
\]
\be
\hat{e}_{\alpha }^{\ m}\,=-\sqrt{2}\,
e^{\frac{3}{4}\phi }\,e_{\alpha }^{\ \mu
}\,A_{\mu }^{{\rm a}}\,\left( U^{-1}\right) _{{\rm a}}^{\ m}\,,\
\ \ \ \ \hat{e%
}_{{\rm a}}^{\ m}=
e^{-\frac{1}{4}\phi }\,\left( U^{-1}\right) _{{\rm a}}^{\ m}.
\ee
The metric components are then given by:
\be
\hat{\g}_{\mu\nu}=e^{-\frac{3}{2}\phi}\, \g_{\mu\nu}-
2\, e^{\frac{1}{2}\phi}\, A^{\rm a}_{\mu}A^{\rm a}_{\nu}\, ,\ \ \
\hat{\g}_{\mu m}=\sqrt{2}\, e^{\frac{1}{2}\phi}\,
A^{\rm a}_{\mu}U^{\rm a}_{\ m}\,  ,\ \ \
\hat{\g}_{mn}=- e^{\frac{1}{2}\phi}\, U^{\rm a}_{\ m}U^{\rm a}_{\ n}\, ,\label{met}
\ee
similarly for $\hat{\g}^{\mu\nu}$. Using these expressions  for the
gravitational and dilaton terms in the action (\ref{tenaction}) one
gets
\begin{equation}
S_{\hat{G}}+S_{\hat{\phi}}=\Omega \int e\,\left( -\frac{1}{4}\,R+\frac{1}{2}%
\,\partial _{\mu }\phi \,\partial ^{\mu }\phi -\frac{1}{8}\,e^{2\phi
}\,F_{\mu \nu }^{{\rm a}}F^{{\rm a\mu \nu
}}+\frac{1}{32}\,e^{-2\phi }\,f_{{\rm abc}}^{2}\right)
\,d^{4}x,\label{md}
\end{equation}
where
\begin{equation}
F_{\mu \nu }^{{\rm a}}=\partial _{\mu }A_{\nu }^{{\rm a}}-\partial
_{\nu }A_{\mu }^{{\rm a}}+f_{{\rm abc}}A_{\mu }^{{\rm b}}A_{\nu
}^{{\rm c}}\, .
\end{equation}
{\bf 2. The two-form.--} Now, the important role is played by the
antisymmetric tensor field. The corresponding ansatz is
\begin{equation}
\hat{B}_{\mu \nu }=B_{\mu \nu }\,,\ \ \ \ \hat{B}_{\mu m}=-\frac{1}{\sqrt{2}}%
\,A_{\mu }^{{\rm a}}\,U_{m}^{{\rm a}}\,,\ \ \ \ \hat{B}_{mn}=\tilde{B}_{mn},
\end{equation}
where $B_{\mu \nu }=B_{\mu \nu }(x)$, while $\tilde{B}_{mn}$ depend
only on $ z$. Computation of the field strength gives
\[
\hat{H}_{\mu \nu \rho }=H_{\mu \nu \rho }\equiv \partial _{\mu }B_{\nu \rho
}+\partial _{\nu }B_{\rho \mu }+\partial _{\rho }B_{\mu \nu }\,,
\qquad
\hat{H}_{\mu \nu m}=-\frac{1}{\sqrt{2}}\,\left( \partial _{\mu }A_{\nu }^{%
{\rm a}}-\partial _{\nu }A_{\mu }^{{\rm a}}\right) \,U_{\ m}^{{\rm a}}\,,
\]
\be
\hat{H}_{\mu mn}=\frac{1}{2}\,f_{{\rm abc}}\,A_{\mu }^{{\rm a}}\,
U_{\ m}^{{\rm %
b}}\,U_{\ n}^{{\rm c}}\,,
\qquad
\hat{H}_{mnp}=\partial _{m}\tilde{B}_{np}+\partial _{n}\tilde{B}%
_{pm}+\partial _{p}\tilde{B}_{mn}.
\ee
We also require that $
\hat{H}_{mnp}=\frac{1}{2\sqrt{2}}\,f_{{\rm abc}}\,
U_{\ m}^{{\rm a}}\,U_{\ n}^{ {\rm b}}\,U_{\ p}^{{\rm c}}\,.$

The next step is to compute the vielbein projections of $H_{MNP}$.
The result is
\[
\hat{H}_{\alpha \beta \gamma }=e^{\frac{9}{4}\,\phi }\,\left( H_{\alpha
\beta \gamma }-\omega _{\alpha \beta \gamma }\right) ,\ \ \ \ \ \hat{H}%
_{\alpha \beta {\rm a}}=-\frac{1}{\sqrt{2}}\,
e^{\frac{5}{4}\,\phi }\,F_{\alpha
\beta }^{{\rm a}}\,,
\]
\be
\hat{H}_{\alpha {\rm ab}}=0\,,\ \ \ \ \ \hat{H}_{{\rm abc}}
=\frac{1}{2\sqrt{2%
}}\, e^{-\frac{3}{4}\,\phi }\,\,f_{{\rm abc}}\,,
\ee
where $F^{\rm a}_{\alpha\beta}=e_{\alpha}^{\ \mu}
e_{\beta}^{\ \nu}F^{\rm a}_{\mu\nu}$
are the tetrad projections of the gauge field tensor, and
$\omega _{\alpha \beta \gamma }$ are the tetrad projections of the
gauge field Chern-Simons 3-form
\be
\omega _{\mu \nu \rho }=-6\left( A_{[\mu }^{{\rm a}}
\partial _{\nu }A_{\rho
]}^{{\rm a}}+\frac{1}{3}\,f_{{\rm abc}}\,A_{\mu }^{{\rm a}}\,
A_{\nu }^{{\rm b%
}}\,A_{\rho }^{{\rm c}}\right) .
\ee
It is now straightforward to compute the last term in the action
(\ref{tenaction}):
\begin{equation}
S_{\hat{F}}=\Omega \int e\,\left( -\frac{1}{8}\,e^{2\phi }\,F_{\mu \nu }^{%
{\rm a}}F^{{\rm a\mu \nu }}-\frac{1}{96}\,e^{-2\phi }\,f_{{\rm abc}}^{\,2}+%
\frac{1}{12}\, e^{4\phi }\, H_{\mu \nu \rho }^{\prime }H^{\prime }{}^{\mu \nu \rho
}\right) \,d^{4}x,
\end{equation}
where $ H_{\mu \nu \rho }^{\prime }=H_{\mu \nu \rho }-\omega _{\mu
\nu \rho }\,.$
Now, taking advantage of the identity $
\varepsilon ^{\sigma \mu \nu \rho }\,\partial _{\sigma }H_{\mu \nu \rho }=0$,
it is easy to see that the expression $
-\Omega \int \left( \frac{1}{6}\,\varepsilon ^{\sigma \mu \nu \rho
}\,\partial _{\sigma }{\bf a}\,H_{\mu \nu \rho }\right) d^{4}x $
vanishes up to a surface term; here ${\bf a}$ is a Lagrange
multiplier. Adding this to the action (\ref{md}) it is possible to
go to a first order formalism where both  $H_{\mu \nu \rho }$ and
${\bf a}$ are treated as independent fields. The equation of motion
of ${\bf a}$  implies that $H_{\mu \nu \rho }$ is a closed form and
can be expressed locally as the curl of $B_{\mu \nu }$. This is
equivalent to varying $H_{\mu
\nu \rho }$ in the action with the result
\begin{equation}
H_{\mu \nu \rho }=\omega _{\mu \nu \rho }+e^{-4\phi }\varepsilon _{\sigma
\mu \nu \rho }\,\partial ^{\sigma }{\bf a}\, .
\end{equation}
Combining the above results gives
\[
S_{10}=\Omega \int e\,\left( -\frac{1}{4}\,R+\frac{1}{2}\,\partial
_{\mu }\phi \,\partial ^{\mu }\phi +\frac{1}{2}\,e^{-4\phi
}\,\partial _{\mu }{\bf 
a}-\frac{1}{4}\,e^{2\phi }\,F_{\mu \nu }^{{\rm a}}F^{{\rm a\mu
\nu }}\right.
\]
\begin{equation}
\left. -\frac{1}{2}\,{\bf a}\,F_{\mu \nu }^{{\rm a}}*\!F^{{\rm a\mu \nu }}+%
\frac{1}{48}\,e^{-2\phi }\,f_{{\rm abc}}^{\,2}\right) \,\,d^{4}x. \label{FS}
\end{equation}
Finally, choosing $U_{\ m}^{{\rm a}}$ and $f_{{\rm abc}}$ for the
$SU(2)\times SU(2)$ case gives $\left( f_{ {\rm abc}}\right)
^{2}=6\,(g_{1}^{2}+g_{2}^{2})$, and thus the dimensionally reduced
action exactly reproduces the bosonic part of the Friedman-Schwarz
model of N=4 supergravity -- up to an overall factor. Analysis of
the  fermionic sector is done in \cite{CV}.

\section{BPS Monopole solutions of N=4 gauged supergravity}
For a purely bosonic configuration, the supersymmetry
transformation laws of the action (\ref{FS}) are \cite{FS} :
\[
\delta \bar{\chi}=-\frac{i}{\sqrt{2}}\,\bar{\epsilon}\,\gamma ^{\mu
}\partial _{\mu }\phi -\frac{1}{2}e^{\phi }\,\bar{\epsilon}\,\alpha
^{a}F_{\mu \nu }^{a}\,\sigma ^{\mu \nu }+\frac{1}{4}\,e^{-\phi }\,\bar{
\epsilon},
\]
\begin{equation}
\delta \bar{\psi}_{\rho }=\bar{\epsilon}
\left( \overleftarrow{\partial}_{\rho
}
-\frac{1}{2}\,\omega _{\rho mn}\,\sigma ^{mn}+\frac{1}{2}\,\alpha
^{a}A_{\rho }^{a}\right) -\frac{1}{2\sqrt{2}}\,e^{\phi }\,\bar{\epsilon}%
\,\alpha ^{a}F_{\mu \nu }^{a}\,\gamma _{\rho }\,
\sigma ^{\mu \nu }+\frac{i}{4
\sqrt{2}}\,e^{-\phi }\,\bar{\epsilon}\,\gamma _{\rho },  \label{st}
\end{equation}
the variations of the bosonic fields being zero. In these formulas, $
\epsilon \equiv \epsilon ^{\rm{I}}$ are four Majorana spinor
supersymmetry parameters, $\alpha ^{a}\equiv \alpha _{\rm{IJ}}^{a}$ are
the SU(2) gauge group generators, whose explicit form is given in \cite{FS},
and $\omega _{\rho mn}$ is the tetrad connection.

We shall consider static, spherically symmetric, purely magnetic
configurations of the bosonic fields, and for this we parameterize
the fields as follows \cite{BM}:
\[
ds^{2}=N\sigma ^{2}dt^{2}-\frac{dr^{2}}{N}-r^{2}(d\theta ^{2}+\sin
^{2}\theta \,d\varphi ^{2}),
\]
\begin{equation}
\alpha ^{a}A_{\mu }^{a}dx^{\mu }=w\ (-\alpha ^{2}\,d\theta +\alpha
^{1}\,\sin \theta \,d\varphi )+\alpha ^{3}\,\cos \theta \,d\varphi ,
\label{metric}
\end{equation}
where $N$, $\sigma $, $w$, as well as the dilaton $\phi $, are
functions of the radial coordinate $r$. Now, since we are unable to
directly solve the equations of motion, we shall consider the
supersymmetry conditions  (\ref{st}), which will give us a set of
first integrals.

The field configuration (\ref{metric}) is supersymmetric, provided
that there are non-trivial supersymmetry Killing spinors $\epsilon$
for which the variations of the fermion fields defined by Eqs.
(\ref{st}) vanish. Inserting configuration (\ref{metric}) into Eqs.
(\ref{st}) and putting $\delta\bar{\chi}=\delta\bar{\psi}
_{\mu}=0$, the supersymmetry constraints become a system of
equations for the four spinors $\epsilon^{\rm{I}}$.

The consistency of the algebraic constraints requires that the determinants
of the corresponding coefficient matrices vanish and that the
matrices commute with each other. These consistency conditions
can be expressed by the  following relations for the background:
\begin{equation}
N\sigma ^{2}=e^{2(\phi -\phi _{0})},  \label{5a}
\end{equation}
\begin{equation}
N=\frac{1+w^{2}}{2}+e^{2\phi }\,\frac{(w^{2}-1)^{2}}{2r^{2}}+\frac{r^{2}}{8}
e^{-2\phi },  \label{5b}
\end{equation}
\begin{equation}
r\phi ^{\prime }=\frac{r^{2}}{8N}e^{-2\phi }\left( 1-4e^{4\phi }\,
\frac{(w^{2}-1)^{2}}{r^{4}}\right) ,  \label{5c}
\end{equation}
\begin{equation}
rw^{\prime }=-2w\frac{r^{2}}{8N}e^{-2\phi }\left( 1+2e^{2\phi }
\frac{w^{2}-1}{r^{2}}\right) ,  \label{5d}
\end{equation}
with constant $\phi _{0}$. Under these conditions, the solution of
the algebraic constraints yields $\epsilon $ in terms of only two
independent functions of $r$. The remaining differential constraint
then uniquely specify these two functions up to two integration
constants, which finally corresponds to two complex unbroken
supersymmetries. We therefore conclude that the supersymmetry
conditions for the bosonic background (\ref{metric}) are given in
terms of Eqs. (\ref{5a})--(\ref{5d}). Solutions of these Bogomolny
equations describe the BPS states in N=4 gauge supergravity with
1/4 of the supersymmetries preserved.

Suppose  that $w(r)$ is not  constant. Introducing the new
variables $ x=w^{2}$ and $R^2=\frac{1}{2}r^{2}e^{-2\phi }$, Eqs.
(\ref{5a})--(\ref{5d}) become equivalent to one first order
differential equation
\begin{equation}
2xR\, (R^2+x-1)\frac{dR}{dx}+(x+1)\,R^2+(x-1)^{2}=0.  \label{6}
\end{equation}
If $R(x)$ is known, the radial dependence of the functions, $x(r)$
and $R(r)$, can be determined from (\ref{5c}) or (\ref{5d}). Eq.
(\ref{6}) is solved by the following substitution:
\begin{equation}
x=\rho^{2}\,e^{\xi (\rho)},\ \ \ \ \ \ \
R^2=-\rho\frac{d\xi (\rho)}{d\rho}-\rho^{2}\,e^{\xi
(\rho)}-1,  \label{7}
\end{equation}
where $\xi (\rho)$ is obtained from
\begin{equation}
\frac{d^{2}\xi (\rho)}{d\rho^{2}}=2\, e^{\xi (\rho)}.  \label{8}
\end{equation}
The most general (up to reparametrizations) solution of this
equation which ensures that $R^2>0$ is $\xi
(\rho)=-2\ln\sinh(\rho-\rho_0)$. This gives us the general solution
of  Eqs. (\ref{5a})--(\ref{5d}). The metric is non-singular at the
origin if only $ \rho_{0}=0 $, in which case
\begin{equation}                    \label{R}
R^{2}(\rho)=2\rho\coth \rho-\frac{\rho^{2}}{\sinh ^{2}\rho}-1,
\end{equation}
one has $R^{2}(\rho)=\rho^{2}+O(\rho^{4})$ as $\rho\rightarrow 0$,
and $R^{2}(\rho)=2\rho+O(1)$ as $\rho\rightarrow \infty $. The last
step is to obtain $r(s)$ from Eq. (\ref{5d}), which finally gives
us a family of completely regular solutions of the Bogomolny
equations:
\begin{equation}
d{ s}^{2}=a^2\,
\frac{\sinh \rho}{R(\rho)}\left\{
dt^{2}-d\rho^{2}-R^{2}(\rho)(d\vartheta ^{2}+\sin ^{2}\vartheta d\varphi
^{2})\right\} ,  \label{10}
\end{equation}
\begin{equation}
w=\pm \frac{\rho}{\sinh \rho},\ \ \ \
e^{2\phi }=a^2\, \frac{\sinh \rho}{2\,R(\rho)},
\label{11}
\end{equation}

\noindent
where $0\leq \rho<\infty $, $R(\rho)$ is given by Eq. (\ref{R}),
and we have chosen in Eq. (\ref{5a}) $2\phi_0=-\ln 2$. The geometry
described by the line element (\ref{10}) is everywhere regular, the
coordinates covering the whole space whose topology is R$^4$. The
geometry becomes flat at the origin, but asymptotically it is not
flat. The shape of the gauge field amplitude $w(\rho)$, given by
Eq. (\ref{11}), corresponds to the gauge field of the regular
magnetic monopole type. In fact, replacing $\rho$ by $r$, the
amplitude exactly coincides with that for the flat space BPS
solution.

\section{Lifting the solution to ten and eleven dimensions}
The results of the previous section imply that any solution of the
gauged supergravity model in four dimensions given in terms of the
metric $\g_{\mu\nu}$, gauge fields $A^{(\C)\, a}_{\mu}$, the axion
$\A$ and the dilaton $\phi$, can be lifted to ten dimensions as a
solution of the N=1 supergravity. The ten-dimensional metric, the
vielbein and the dilaton $\hat{\phi}$ are then given by Eqs.
(\ref{vier}) -- (\ref{met}).

 Choosing $A^{(2)\, a}_{\mu}=g_2=0$ and $g_1
=1$, the lifted solutions can be represented as follows. The metric
and the dilaton are
\be
\hat{\g}_{MN}=2 e^{-\hat{\phi}}\, \tilde{\g}_{MN},\ \ \ \ \ \ \
\hat{\phi}=-\frac{\phi(\rho)}{2}\, ,\label{sm}
\ee
where the metric in the string frame,
$\tilde{\g}_{MN}$, is specified by the line element
\be
d\tilde{s}^2 =
dt^{2}-d\rho^{2}-
R^{2}(\rho)\, d\Omega^{2}_2
-\Theta^a\Theta^a
-(dz^4)^2-(dz^5)^2-(dz^6)^2\, .
\ee
Here $d\Omega^{2}_2$ is the standard metric on the unit 2-sphere,
\be
\Theta^a\equiv A^a- \theta^a=
A^{a}_{\mu}\, dx^\mu - \theta^{a}_{\ i}\, dz^i\, ,
\ee
where $\theta^a$ are the Maurer-Cartan forms on $S^3$ parametrized
by $\{z^i\}=\{z^1,z^2,z^3\}$. If $\T_a$ are the SU(2) group
generators, $[\T_a,\T_b]=i\epsilon_{abc}\T_c\, $, then the gauge
field  is given by
\be                                                \label{6.2:2}
A\equiv\T_a A^{a}\equiv
\T_{a}A_{ \mu }^{a}dx^{\mu }=
w(\rho)\,
\{-\T_{2}\,d\theta +\T
_{1}\,\sin \theta \,d\varphi \}+\T_{3}\,\cos \theta \,d\varphi\, .
\ee
The non-vanishing vielbein projections
of the antisymmetric tensor field are
\be
\hat{H}_{\alpha\beta a}=
-\frac{1}{2\sqrt{2}}\, e^{-\frac{3}{4}\phi}\, F^{a}_{\alpha\beta},\ \ \ \
\hat{H}_{abc}=
\frac{1}{2\sqrt{2}}\, e^{-\frac{3}{4}\phi}\, \epsilon_{abc}\, ,
\ee
where $F^{a}_{\alpha\beta}$ are the tertad projections of the gauge
field tensor corresponding to the gauge field (\ref{6.2:2}) for the
tetrad $e_\alpha$ specified by the four-dimensional part of the
string metric (\ref{sm}).

To lift the solution further to eleven dimensions we first note
that N=1 supergravity in ten-dimensions is obtained from eleven
dimensional supergravity by dimensional reduction and then
compactification. One has the following identifications:
$e_M^A=e^{-\frac{1}{6}\hat{\phi}}\hat{e}_M^A $ and
$e_{\stackrel{.}{11}}^{11}=e^{\frac{4}{3}\hat{\phi}}$. For the
antisymmetric tensor we have $A_{MN\stackrel{.}{11}}=B_{MN}$ and
$A_{MNP}=0$. This implies that we can write the different
components of the eleven dimensional metric as follows:

\[
g_{\mu \nu }^{(11)}=e^{-\frac{4}{3}\phi }\left( g_{\mu \nu
}-2e^{2\phi }A_{\mu }^{a}A_{\nu }^{a}\right)  \qquad g_{\mu
m}^{(11)} =\sqrt{2}e^{\frac{2}{3}\phi }A_{\mu }^{a}U_{m}^{a}
\]
\be
g_{mn}^{(11)} =-e^{\frac{2}{3}\phi }U_{m}^{a}U_{n}^{a} \qquad
g_{\stackrel{.}{11}\stackrel{.}{11}}^{(11)} =e^{-\frac{4}{3}\phi }.
\ee

\end{document}